\documentclass[journal,draftclsnofoot,onecolumn,12pt]{IEEEtran}

\usepackage{amsmath}
\usepackage{amssymb}
\usepackage[dvips]{graphicx}
\graphicspath{{./fig/}}
\usepackage{curves,color}
\usepackage{subfigure}
\usepackage{dsfont,slashbox,wasysym}
\usepackage{tikz}
\usepackage{pdfpages,enumerate}
\usepackage{multirow}
\usepackage{bm}
\usepackage{algorithmic}

\allowdisplaybreaks

\newtheorem{theorem}{Theorem}
\newtheorem{lemma}{Lemma}

\newtheorem{example}{Example}
\newtheorem{definition}{Definition}
\newtheorem{proposition}{Proposition}
\newtheorem{remark}{Remark}
\newtheorem{observation}{Observation}
\newtheorem{cor}{Corollary}
\newtheorem{claim}{Claim}

\RequirePackage[normalem]{ulem} 
\RequirePackage{color}\definecolor{RED}{rgb}{1,0,0}\definecolor{BLUE}{rgb}{0,0,1} 

\begin{document}
\title{Linear Index Coding With Multiple Senders and Extension to a Cellular Network}
\author{Jae-Won~Kim~and~Jong-Seon~No, \IEEEmembership{Fellow,~IEEE}
\thanks{This work was supported by the National Research Foundation of Korea (NRF) grant funded by the Korea government (MSIP) (No. NRF-2016R1A2B2012960). 

J.-W.~Kim and J.-S.~No are with the Department of Electrical and Computer Engineering, INMC, Seoul National University, Seoul 08826, Korea (e-mail: kjw702@ccl.snu.ac.kr, jsno@snu.ac.kr).

}
}

\maketitle

\begin{abstract}

In this paper, linear index codes with multiple senders are studied, where every receiver receives encoded messages from all senders. A new fitting matrix for the multiple senders is proposed and it is proved that the minimum rank of the proposed fitting matrices is the optimal codelength of linear index codes for the multiple senders. In addition, a new type of a side information graph related with the optimal codelength is proposed and whether given side information is critical or not is studied. Furthermore, linear index codes for the cellular network scenario are studied, where each receiver can receive a subset of sub-codewords. Since some receivers cannot receive the entire codeword in the cellular network scenario, the encoding method based on the fitting matrix has to be modified. In the cellular network scenario, we propose another fitting matrix and prove that an optimal generator matrix can be found based on these fitting matrices. In addition, some properties on the optimal codelength of linear index codes for the cellular network case are studied. 
\end{abstract}

\vspace{2pt}
\section{Introduction}\label{sec:introduction}
\IEEEPARstart{I}{ndex} coding was first introduced by Birk and Kol \cite{Informed} to utilize side information of each receiver in the error-free broadcast channel. In order to find the optimal index code, a lot of index coding schemes have been researched \cite{ICSI}--\cite{codingscheme4}.

\begin{figure}[t]
\centering
\subfigure[]{\includegraphics[scale=0.45]{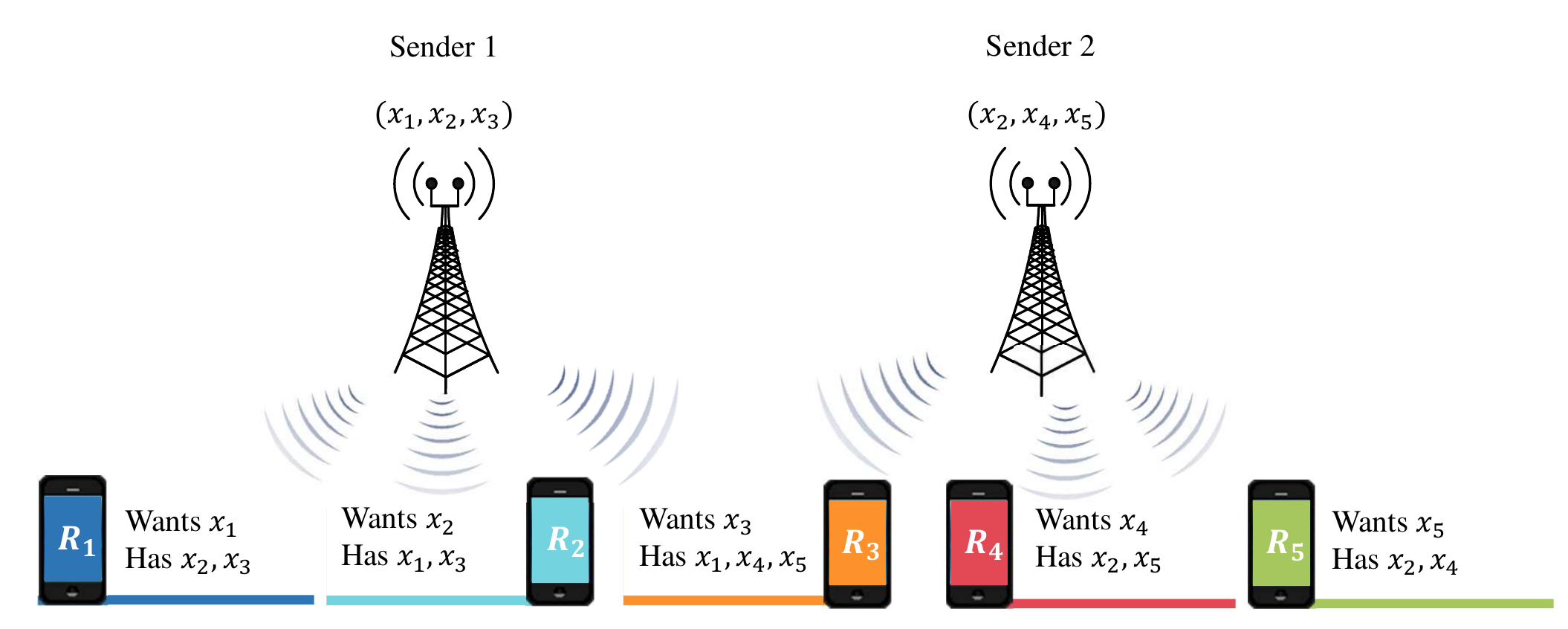}
\label{fig:cellularex}}
\subfigure[]{\includegraphics[scale=0.5]{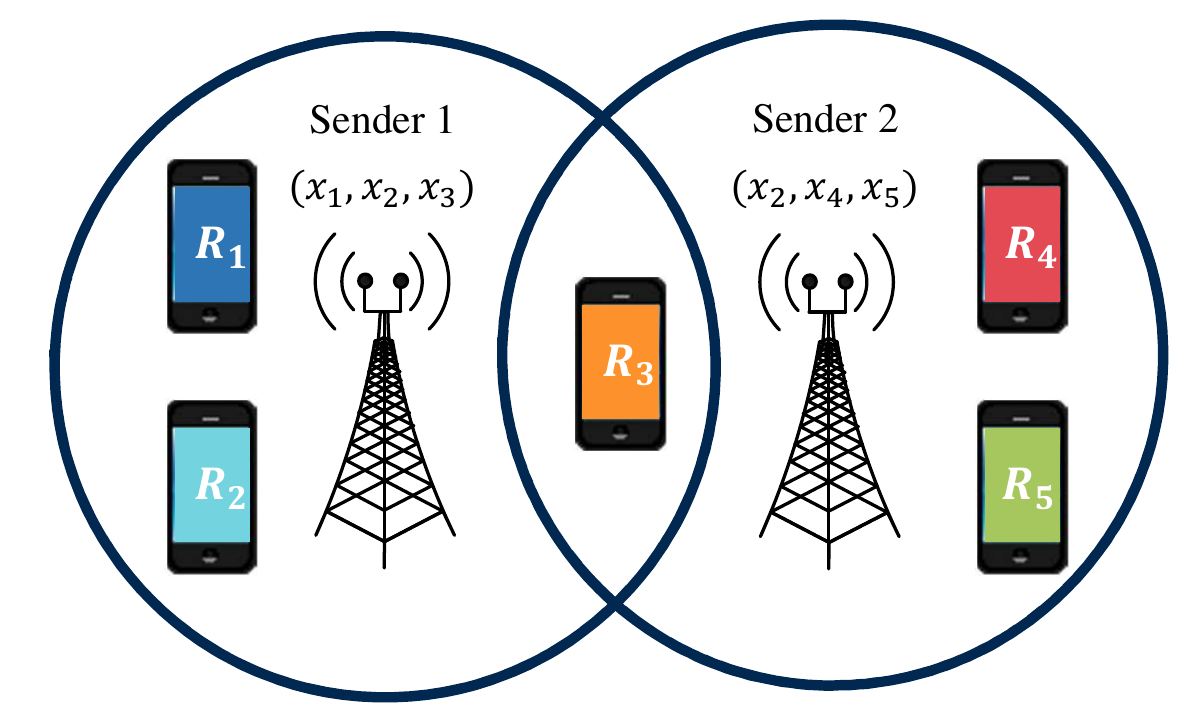}
\label{fig:cellularcover}}
\caption{A description for index coding with two senders in a cellular network: (a) An index coding instance. (b) Coverage of two senders.} 
\label{fig:cellularexplanation}
\end{figure}

In addition to researches on finding index coding schemes, there are some researches on finding the relationship between index coding and other problems. In \cite{EQU}, it was proved that any network coding instance can be transformed to the corresponding index coding instance and a solution for the network coding instance exists if and only if a solution for the corresponding index coding instance exists. It was studied that topological interference management can be performed using index coding \cite{TIM}. Furthermore, it was shown that there is a duality between distributed storage and index coding \cite{DSS}, \cite{DSS2}.

To deal with the more realistic index coding instance, index codes with side information errors were researched \cite{ICSIE} considering memory errors and index coding with multiple senders was introduced \cite{MSIC}. Since there are a lot of scenarios where messages are distributed among multiple senders, index coding with multiple senders has attracted significant attention and a lot of researches on multiple senders have been done to find the capacity region. Graph-theoretic approaches to the two sender index coding problem were researched \cite{TSIC}. In \cite{SMSIC}, partitioned distributed composite coding was studied based on composite coding in \cite{RI} for a general multiple sender case. In \cite{CMSIC}, multi-sender cooperative composite coding was developed and advantages of cooperative composite coding were studied.

Existing researches on index coding with multiple senders consider a scenario, where every receiver can receive encoded messages from all senders. That assumption is valid if every receiver belongs to coverage of each sender. However, some receivers do not belong to coverage of all senders in reality. For example, some receivers can receive encoded messages from a subset of senders in a cellular network due to the problem of coverage. Thus, index coding in a cellular network has to be studied to utilize index coding in the realistic scenario. In a cellular network, an index coding instance where multiple senders exist and receivers are restricted to receive encoded messages from some senders has to be considered and this scenario for two senders is depicted in Fig. \ref{fig:cellularexplanation}. 

In Fig. \ref{fig:cellularexplanation}, receivers 1 and 2 can receive encoded messages from only sender 1, receivers 4 and 5 can receive encoded messages from only sender 2, and receiver 3 can receive all encoded messages. Then, every receiver is satisfied if sender 1 transmits $(x_1+x_2+x_3)$ and sender 2 transmits $(x_2+x_4+x_5)$. Since receiver 3 can receive all encoded messages, receiver 3 can calculate $(x_1+x_2+x_3)-(x_2+x_4+x_5)=x_1+x_3-x_4-x_5$. By using its side information, receiver 3 can recover $x_3$.

In this paper, we study some properties on linear index codes with multiple senders. Since there are multiple senders, a lot of properties of index codes with the single sender have to be modified. First, a fitting matrix for multiple senders is introduced and an encoding method using the fitting matrix is also studied as in \cite{ICSI}. Then, some properties related with the optimal codelength are studied and whether given side information is critical or not is studied for multiple senders. 

Furthermore, linear index coding with multiple senders in a cellular network is studied. In a cellular network with two senders, another type of fitting matrices is introduced and the encoding method based on the fitting matrix is studied to find the optimal codelength. In addition, some properties on the optimal codelength of linear index codes for the cellular network case are studied. 

This paper is organized as follows. In Section \ref{sec:preliminary}, the problem setup is introduced and the encoding method of linear index codes based on the fitting matrix is studied in Section \ref{sec:Encoding}. Next, some properties of linear index codes with multiple senders are studied in Section \ref{sec:Properties}. Then, index coding in a cellular network is studied in Section \ref{sec:cellularnetwork}. Finally, conclusions are given in Section \ref{Sec:Conclusion}.

\vspace{2pt}
\section{Preliminary}\label{sec:preliminary}

In this paper, linear index codes with multiple senders are considered. We first introduce some notations and then describe a linear index coding problem for the multiple senders.

\subsection{Notations}
Let $\mathbb{F}_q$ be the finite field of size $q$ and $Z[n]=\{1,2,\ldots,n\}$ for a positive integer $n$. For a vector $\bold{x}\in \mathbb{F}_q^n$, $\rm{\rm{wt}}(\bold{x})$ denotes the Hamming weight of $\bold{x}$. Let $\bold{x}_A$ be a subvector $(x_{i_1},x_{i_2},\ldots,x_{i_{|A|}})$ of a vector $\bold{x}=(x_1,x_2,\ldots,x_n)\in \mathbb{F}_q^n$ for a subset $A=\{i_1,i_2,\ldots,i_{|A|}\}\subseteq Z[n]$, where $i_1<i_2<\ldots<i_{|A|}$.

\subsection{Problem Formulation}
 
In the linear index coding problem for the case of multiple senders, there are $|S|$ senders and $m$ receivers $R_1,\ldots,R_m$, where $S=\{s_1,s_2,\ldots,s_{|S|}\}$ is a set of senders. There are $n$ messages represented by $\bold{x}=(x_1,\ldots,x_n)\in\mathbb{F}_q^n$, which are distributed in $|S|$ senders so that $\cup_{i\in Z[|S|]} M_i=Z[n]$, where $M_i$ denotes a set of indices of messages which $s_i$ has for $i\in Z[|S|]$. Let $M=\{M_1,\ldots,M_{|S|}\}$ and $M_c=\{i~|~\exists s_j, s_k {\rm ~such~ that~} i\in M_j {\rm ~and~} i\in M_k\}$ for any $j,k\in Z[|S|]$ such that $j\neq k$.

Each sender broadcasts its encoded messages through the error-free broadcast channel. Each receiver $R_i$ has $\bold{x}_{\mathcal{X}_i}$ as side information and wants to receive the wanted message denoted by $x_{f(i)}$, where $f(i)$ is an index of the message that $R_i$ wants to receive and $\mathcal{X}_i$ is a set of side information indices of $R_i$. Each receiver $R_i$ receives all encoded messages from $|S|$ senders and $R_i$ has to recover $x_{f(i)}$ from the received codeword and $\bold{x}_{\mathcal{X}_i}$. We assume that $\{f(i)\}\cap\mathcal{X}_i=\phi$ and let $\mathcal{Y}_i=Z[n]\setminus\{f(i)\}\setminus\mathcal{X}_i$ for $i\in Z[m]$.

Let $\mathcal{G}$ be a bipartite side information graph, where a directed edge from a receiver node to a message node means that the receiver has the message as side information and a directed edge from a message node to a receiver node means that the receiver wants to receive the message. It is assumed that $s_1$ knows $\mathcal{G}$ and $M$, $s_1$ determines encoding procedures of all senders, and each sender can receive its encoding strategy from $s_1$. 

Then, a linear index code for multiple senders is defined as follows.

\vspace{2mm}
\begin{definition}
A linear index code with multiple senders over $\mathbb{F}_q$, denoted by a $(\mathcal{G},M)$-IC is a set of codewords having:
\begin{enumerate}
\item Generator submatrices $G^{(i)}\in\mathbb{F}_q^{|M_i|\times N_i}$ for $i\in Z[|S|]$, where $N_i$ denotes the codelength of the sub-codeword $\bold{x}_{M_i}G^{(i)}$ generated by $s_i$.
\item Decoding functions $D_j : \mathbb{F}_q^{N_1+N_2+\cdots+N_{|S|}}\times\mathbb{F}_q^{|\mathcal{X}_j|}\rightarrow\mathbb{F}_q$ satisfying 
\begin{equation}
D_j(\bold{x}_{M_1}G^{(1)},\ldots,\bold{x}_{M_{|S|}}G^{(|S|)},\bold{x}_{\mathcal{X}_j})=x_{f(j)}
\nonumber
\end{equation}
for all $j\in Z[m]$, $\bold{x}\in\mathbb{F}_q^n$.
\end{enumerate}
\end{definition}
\vspace{2mm}

It is noted that an $n\times N$ generator matrix $G$ for a $(\mathcal{G},M)$-IC is constructed by using generator submatrices $G^{(i)}$ for $i\in Z[|S|]$ and the codelength $N$ is $N_1+\cdots+N_{|S|}$. Let $N_{\rm opt}^q(\mathcal{G},M)$ be the optimal codelength of a $(\mathcal{G},M)$-IC. Since there are more than one sender, another graph representing them is needed as in the following definition \cite{MSIC}.

\vspace{2mm}
\begin{definition}
A message graph $U$ is a unipartite graph of $n$ message nodes, where an undirected edge between any two nodes $i$ and $j$ exists if and only if $x_i$ and $x_j$ are known to the same sender, that is, $i,j\in M_s$ for some $s\in Z[|S|]$ such that $i\neq j$.
\end{definition}
\vspace{2mm}

\begin{figure}[t] 
\centering
\includegraphics[scale=0.6]{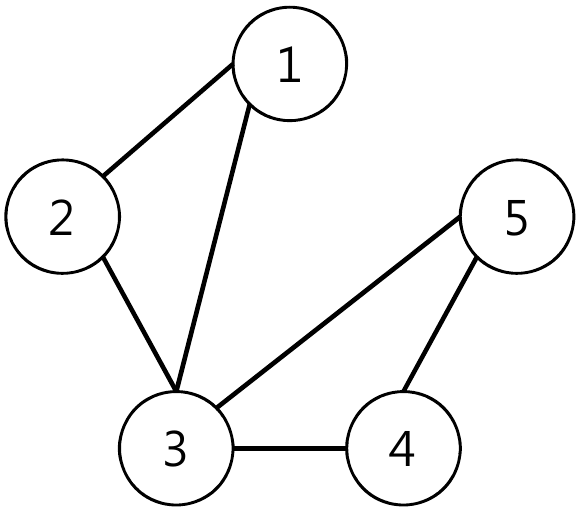}
\caption{A message graph $U$ for $n=5$, $M_1=\{1,2,3\}$, and $M_2=\{3,4,5\}$.} 
\label{fig:messagegraph}
\end{figure}

From a message graph $U$, it is noted that which two messages are contained in the same sender, that is, which two messages can be encoded together. In general, there is the constraint of multiple senders that two messages not connected in $U$ cannot be encoded together. Fig. \ref{fig:messagegraph} is an example of $U$ for $n=5$, $M_1=\{1,2,3\}$, and $M_2=\{3,4,5\}$. 

%

We have a well known claim of linear index codes as follows \cite{ICSI}.

\vspace{2mm}
\begin{claim}
If $G$ is a generator matrix of a linear index code, $\Sigma_{i\in Z[N]} a_i^kG_i=e_{f(k)}+\Sigma_{j\in\mathcal{X}_k}b_j^ke_j$ for some $a_i^k, b_j^k\in\mathbb{F}_q$, where $k\in Z[m]$, $G_i$ denotes the $i$th column of $G$, and $e_j$ denotes the $j$th standard basis vector.

\label{claim:ICSI}
\end{claim}

\vspace{2pt}
\section{Encoding Procedure of $(\mathcal{G},M)$-IC}\label{sec:Encoding}

In this section, we first introduce a new type of a fitting matrix for the multiple senders. Then, it is proved that the minimum rank of the proposed fitting matrices is the same as $N_{\rm opt}^q(\mathcal{G},M)$. In \cite{CMSIC2}, they proposed a method to find the optimal codelength of linear index codes. They suggested many strategies for search space reduction and checked decoding conditions of receivers for every generator matrix candidate. Unlike their methods, we consider the fitting matrix which always satisfies decoding conditions of receivers and propose a new method to find the optimal codelength based on the fitting matrix. The fitting matrix for the multiple senders consists of $|S|$ submatrices as shown in the following definition.

\vspace{2mm}
\begin{definition}
The fitting matrix $F$ of size $n\times m|S|$ for a $(\mathcal{G},M)$-IC is described as follows:
\begin{enumerate}
\item There are $|S|$ submatrices $F^{(1)},F^{(2)},\ldots,F^{(|S|)}$ of size $n\times m$ representing each sender.
\item $F_{i,k}^{(j)}=0$ for $i\notin M_j$, $j\in Z[|S|]$, and $k\in Z[m]$.
\item $\Sigma_{j=1}^{|S|}a_j^kF_{f(k),k}^{(j)}=1$ for $a_j^k\in\mathbb{F}_q$ and $k\in Z[m]$.
\item $\Sigma_{j=1}^{|S|}a_j^kF_{i_1,k}^{(j)}=0$ for the same $a_j^k\in\mathbb{F}_q$ as the above, $i_1\in\mathcal{Y}_k$, and $k\in Z[m]$.
\item For $i_2\in M_j\cap\mathcal{X}_k$, $j\in Z[|S|]$, and $k\in Z[m]$, $F_{i_2,k}^{(j)}$ is any element of $\mathbb{F}_q$.

\end{enumerate}
\label{def:fittingmatrix}
\end{definition}
\vspace{2mm}

For linear index codes, the minimum rank of the conventional fitting matrices for the single sender IC problem is known to be the optimal codelength \cite{ICSI}. Similarly, the minimum rank of the fitting matrices for a $(\mathcal{G},M)$-IC is the optimal codelength as shown in the following theorem.

\vspace{2mm}
\begin{theorem}
Let $F^{\prime}$ be a fitting matrix having the minimum rank. Then, an optimal generator matrix of a $(\mathcal{G},M)$-IC is the matrix $G^\prime$ generated by deleting linearly dependent columns in $F^{\prime}$.
\begin{IEEEproof}
First, we show that a fitting matrix of a $(\mathcal{G},M)$-IC can be a generator matrix. From 2) of Definition \ref{def:fittingmatrix}, the index code corresponding to the fitting matrix can be made for multiple senders. Then, for $k\in Z[m]$, $R_k$ can obtain $x_{f(k)}$ by using its side information and the linear combination of the $k$th columns of submatrices represented in Definition \ref{def:fittingmatrix}, that is, the linear combination of the received codeword components. 

Next, we show that the minimum rank of these fitting matrices is the optimal codelength. Let $G$ be an $n\times N$ generator matrix of a $(\mathcal{G},M)$-IC. From Claim \ref{claim:ICSI}, it is noted that $\Sigma_{i\in Z[N]} a_i^kG_i=e_{f(k)}+\Sigma_{j\in\mathcal{X}_k}b_j^ke_j$ for some $a_i^k, b_j^k\in\mathbb{F}_q$, where $k\in Z[m]$ and $G_i$ denotes the $i$th column of $G$.

By the definition of the fitting matrix $F$, it is noted that columns whose linear combination is $e_{f(k)}+\Sigma_{j\in\mathcal{X}_k}b_j^ke_j$ can be the $k$th columns of submatrices of $F$ if each column satisfies 2) of Definition \ref{def:fittingmatrix}. Since every column $G_i$ is used for the encoding in one sender, we can classify which column of $G$ belongs to which sender. Let $C_l=\{i|i\in Z[N] {\rm~and~} G_i {\rm~is~used~for~encoding~in~} s_l\}$ for $l\in Z[|S|]$. Then, we can split $\Sigma_{i\in Z[N]} a_i^kG_i$ into $\Sigma_{i\in C_1} a_i^kG_i+\cdots+\Sigma_{i\in C_{|S|}} a_i^kG_i$. If we make the $k$th column of $F^{(l)}$ as $\Sigma_{i\in C_l} a_i^kG_i$ for all $k$ and $l$, it becomes a fitting matrix for a $(\mathcal{G},M)$-IC and it is obvious that the rank of this fitting matrix is smaller than or equal to $N$. Since the rank of a generator matrix is the same as codelength, $G^\prime$ is an optimal generator matrix of a $(\mathcal{G},M)$-IC.
\end{IEEEproof}
\label{theorem:multisenderencoding}
\end{theorem}

Now, we have the following corollary specifying the fitting matrix $F$ for a $(\mathcal{G},M)$-IC.

\vspace{2mm}
\begin{cor}
Coefficients in 3) and 4) of Definition \ref{def:fittingmatrix} can be modified as follows:
\begin{enumerate}
\item $\Sigma_{j=1}^{|S|}F_{f(k),k}^{(j)}=1$ for $k\in Z[m]$.
\item $\Sigma_{j=1}^{|S|}F_{i_1,k}^{(j)}=0$ for $i_1\in\mathcal{Y}_k$ and $k\in Z[m]$.
\end{enumerate}

\begin{IEEEproof}
For the perspective of the minimum rank, nonzero coefficients in 3), 4) of Definition \ref{def:fittingmatrix} do not change the rank of the fitting matrix. For zero coefficients, if we make the corresponding columns as all-zero columns, the rank of the modified fitting matrix becomes smaller than or equal to that of the original one and it is also a fitting matrix for a $(\mathcal{G},M)$-IC. 
\end{IEEEproof}
\label{cor:reduce}
\end{cor}

\begin{figure}[t]
\centering
\includegraphics[scale=0.6]{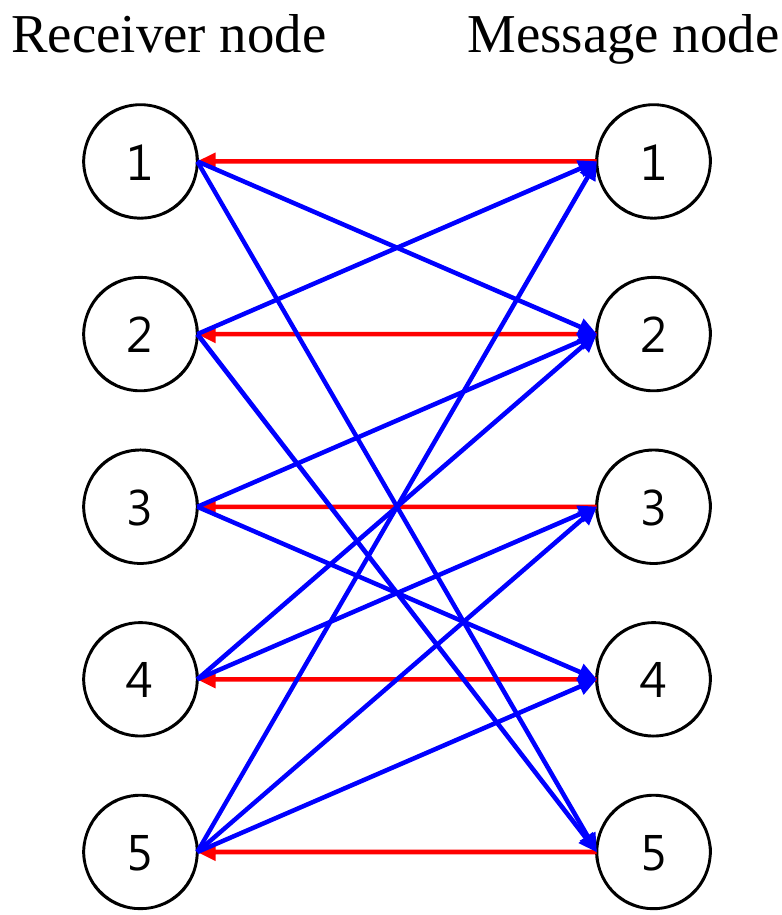}
\caption{A side information graph $\mathcal{G}$ for $m=n=5$.} 
\label{fig:figfitting}
\end{figure}

We have the following remark for the encoding procedure based on the fitting matrix.

\vspace{2mm}
\begin{remark}
It is noted that the sum of the $i$th column of each submatrix of the fitting matrix $F$ for a $(\mathcal{G},M)$-IC is the transpose of the $i$th row of the conventional fitting matrix for the single sender IC problem with $\mathcal{G}$ \cite{ICSI}, where $i\in Z[m]$. After choosing linearly independent columns of $F$, it is noted that selected columns in $F^{(j)}$ are used for the encoding procedure of $s_j$ for $j\in Z[|S|]$.
\label{rmk:fitting}
\end{remark}
\vspace{2mm}

From Remark \ref{rmk:fitting}, it is noted that we can construct a fitting matrix systematically from Definition \ref{def:fittingmatrix}. By considering coefficients of 3) and 4) in Definition \ref{def:fittingmatrix} as $1$s, we can determine all elements of $F$ as in the following example.

\vspace{2mm}
\begin{example}
Let $m=n=5$, $M=\{M_1,M_2\}$, $M_1=\{1,2,3\}$, and $M_2=\{3,4,5\}$. A side information graph $\mathcal{G}$ is given in Fig. \ref{fig:figfitting}. Then, the fitting matrix of a $(\mathcal{G},M)$-IC is derived as

\begin{equation}
F=\begin{pmatrix}F^{(1)}&F^{(2)}\end{pmatrix},
\nonumber
\end{equation} where

\begin{equation}
F^{(1)}=\begin{pmatrix}1&*&0&0&*\\{*}&1&*&*&0\\a&b&c&*&*\\0&0&0&0&0\\0&0&0&0&0\end{pmatrix},  F^{(2)}=\begin{pmatrix}0&0&0&0&0\\{0}&0&0&0&0\\-a&-b&1-c&*&*\\0&0&*&1&*\\{*}&*&0&0&1\end{pmatrix}
\nonumber
\end{equation}
and ${*}$ denotes any element of $\mathbb{F}_q$ and $a,b,c\in\mathbb{F}_q$. By finding a fitting matrix having the minimum rank, we can find an optimal generator matrix.
\label{ex:fitting}
\end{example}

Since both a method in \cite{CMSIC2} and the proposed method based on the fitting matrix can find the optimal codelength, we have the following remark to compare them.

\vspace{2mm}
\begin{remark}
Consider $\mathcal{P}^\prime$ in \cite{CMSIC2} and assume that $m=n$ and $f(j)=j$ for $j\in Z[m]$. Then, the complexity for solving $\mathcal{P}^\prime$ is $O\Big(q^{\Sigma_{i\in Z[|S|]} {{|M_i|^2+|M_i|}\over{2}}}\times\big(\Sigma_i |M_i|^2 + \Sigma_j \Sigma_i (1+|\mathcal{Y}_j|)^2|M_i|\big)\Big)$ and the complexity for the proposed method is $O\Big(q^{\Sigma_j\big(\Sigma_i (|\mathcal{X}_j\cap M_i|+|(\mathcal{Y}_j\cup \{j\})\cap M_i\cap M_c|)-|(\mathcal{Y}_j\cup \{j\})\cap M_c|\big)}\times |S|n^3\Big)$. It is not easy to compare the above two complexities. However, it is expected that the smaller $|\mathcal{X}_j|$, $|M_c|$, and $|S|$, the lower the complexity for the proposed method based on the fitting matrix. Since the proposed fitting matrix can be used to derive some properties of index codes in the following sections, we consider the encoding procedure based on the fitting matrix in this paper.
\end{remark}
\vspace{2mm}

The following proposition shows that we can further simplify the fitting matrix $F$ for a $(\mathcal{G},M)$-IC if side information of receivers has some properties.

\vspace{2mm}
\begin{proposition}
For a $(\mathcal{G},M)$-IC, if $\mathcal{X}_i\subset M_j$ and $f(i)\in M_j$ for $i\in Z[m]$ and $j\in Z[|S|]$, the $i$th columns in the submatrices except $F^{(j)}$ can be assumed as all-zero columns in the perspective of the optimal codelength and the $i$th column in $F^{(j)}$ is the transpose of the $i$th row of the conventional fitting matrix of the single sender IC problem.
\begin{IEEEproof}
It is noted that the $i$th column of $F^{(j)}$ can be transformed into the transpose of the $i$th row of the conventional fitting matrix by adding the other $i$th columns of submatrices. Since elementary column operations do not change the rank of matrices and the $i$th column of $F^{(j)}$ can be set to the transpose of the $i$th row of the conventional fitting matrix from Definition \ref{def:fittingmatrix}, the $i$th columns in the other submatrices can be assumed as all-zero columns to minimize the rank of the fitting matrix.
\end{IEEEproof}
\label{prop:sideinformation}
\end{proposition}
\vspace{2mm}

Thus, if every receiver satisfies the condition in Proposition \ref{prop:sideinformation}, it can be reduced to the single sender IC problem.

%
%
%
%
%
%

\vspace{2pt}
\section{Properties of $(\mathcal{G},M)$-IC}
\label{sec:Properties}

A $0$-cycle for an IC is a subgraph of $\mathcal{G}$, which is known to be important in an IC problem for the single sender case \cite{ICSIE}. For the multiple sender case, a $0$-cycle is also related with a lot of properties of a $(\mathcal{G},M)$-IC.

Let $\Phi$ be a set of subsets of $Z[n]$ defined by
\begin{equation}
\Phi=\{B\subseteq Z[n]\big||\mathcal{X}_i\cap B|\geq 1 \textrm{ for all } i\in Z[m] \textrm{ s.t. } f(i)\in B\}
\nonumber
\end{equation}
for a side information graph $\mathcal{G}$ of a $(\mathcal{G},M)$-IC. Then, we have the following definition for a $0$-cycle.

\vspace{2mm}
\begin{definition}
For a $(\mathcal{G},M)$-IC, a subgraph $\mathcal{G}^{\prime}$ of $\mathcal{G}$ induced by an element of $\Phi$ is called a $0$-cycle. Then, $\mathcal{G}$ is said to be $0$-acyclic if there is no $0$-cycle in $\mathcal{G}$.
\label{def:delta cycle}
\end{definition}

Now, we define a special type of a $0$-cycle for a $(\mathcal{G},M)$-IC.

\vspace{2mm}
\begin{definition}
Let $U^\prime$ be a subgraph of $U$, which has message nodes corresponding to a $0$-cycle. Then, a $0$-cycle is said to be message-connected if and only if there is at least one path between any two message nodes in $U^\prime$. Otherwise, a $0$-cycle is said to be message-disconnected.
\end{definition}
\vspace{2mm}

There is an example for a $0$-cycle and a message-connected $0$-cycle.

\vspace{2mm}
\begin{example}
Assume that a message graph and $M$ are given in Fig. \ref{fig:messagegraph} and a side information graph is given in Fig. \ref{fig:figfitting}. Then, a subgraph of $\mathcal{G}$ induced by message nodes $\{1,2,3\}$ and receiver nodes $\{1,2,3\}$ is a $0$-cycle because $2\in\mathcal{X}_1$, $1\in\mathcal{X}_2$, and $2\in\mathcal{X}_3$. Furthermore, it is also a message-connected $0$-cycle because a subgraph of $U$ induced by $\{1,2,3\}$ has at least one path between any two message nodes. A subgraph of $\mathcal{G}$ induced by message nodes $\{1,2,5\}$ and receiver nodes $\{1,2,5\}$ is a $0$-cycle but it is not a message-connected $0$-cycle.
\end{example}
\vspace{2mm}

%

It is said that $x_i$ forms a $0$-cycle if there exists at least one $0$-cycle containing $x_i$ for $i\in Z[n]$, otherwise it is said that $x_i$ does not form a $0$-cycle. The similar argument holds for a message-connected $0$-cycle and a message-disconnected $0$-cycle.

In the perspective of the optimal codelength, some side information does not help to reduce the codelength, that is, removing the corresponding edges in $\mathcal{G}$ does not increase the optimal codelength. Then, it is said that the side information or edges are not critical.
In the following lemmas, some properties of a $0$-cycle are given.

\vspace{2mm}
\begin{lemma}
For a message $x_{i_1}$ such that $i_1\in Z[n]$, $x_{i_1}$ is the same as being sent in the uncoded form if $x_{i_1}$ does not form a $0$-cycle.
\begin{IEEEproof}
Assume that $R_1,\ldots,R_k$ want to receive $x_{i_1}$. If $x_{i_1}$ does not form a $0$-cycle in $\mathcal{G}$, there are two cases. The first one is that there exists a receiver (say $R_1$) which does not have any side information. In this case, it is trivially proved.

The other case is that $x_{i_1}$ does not form a $0$-cycle but each $R_i$ for $i\in Z[k]$ has at least one side information symbol. Without loss of generality, this case means that each of the side information symbols of $R_1$ (say one of them is $x_{i_2}$) does not form a $0$-cycle. Then, we can continue the same procedure for $x_{i_2},x_{i_3},\ldots,x_{i_a}$. Then, it results in the fact that a receiver $R_b$ wanting $x_{i_a}$ does not have any side information symbols, which means that $x_{i_a}$ is the same as being sent in the uncoded form and having $x_{i_a}$ as side information is not critical. Thus, it is the same that $R_1$ does not have any side information symbols. 
\end{IEEEproof}
\label{lemma:uncoded}
\end{lemma}

\vspace{2mm}
\begin{lemma}
For a message $x_i$ such that $i\in Z[n]$, $x_i$  is the same as being sent in the uncoded form if $x_i$ does not form a message-connected $0$-cycle.
\begin{IEEEproof}
There are two cases. The first one is that $x_i$ does not form a $0$-cycle and this case is directly proved from Lemma \ref{lemma:uncoded}. The second case is that $x_i$ only forms a message-disconnected $0$-cycle.

Let $T=\{x_i,x_{i_1},x_{i_2},\ldots,x_{i_t}\}$ be a message set represented by the maximum $0$-cycle including $x_i$. Then, every message forming a $0$-cycle is included in $T$. Since the other messages are the same as being sent in the uncoded forms, having those messages as side information is not critical. Then, partition $T$ into subsets of messages $T_1,T_2,\ldots,T_k$ based on existence of a path between any two messages in the message graph induced by $T$. Since the encoded messages related with $T_q$ do not have any information of messages in $T_p$ for $p,q\in Z[k]$ such that $p\neq q$, having messages in $T_q$ as side information does not help to recover messages in $T_p$. Thus, receivers wanting messages in $T_p$ do not need to know messages in $T_q$ as side information. 


Assume that $x_i\in T_1$. Then, it can be assumed that all side information of receivers wanting messages in $T_1$ is included in $T_1$. Thus, we can consider a side information graph induced by $T_1$. In this graph, let $T^{(1)}$ be a message set represented by the maximum $0$-cycle and partition $T^{(1)}$ into subsets of messages $T^{(1)}_1,T^{(1)}_2,\ldots,T^{(1)}_{k_1}$ based on existence of a path between any two messages in the message graph induced by $T^{(1)}$. If $x_i\notin T^{(1)}$, it is proved from Lemma \ref{lemma:uncoded}. If $x_i\in T^{(1)}$, we assume that $x_i\in T^{(1)}_1$ and continue the same procedure as the above. Then, $x_i$ does not belong to $T^{(j)}$ for some $j$ or $x_i$ belongs to $T^{(h)}$ such that $k_h=1$ or $T^{(h)}$ is $\phi$. However, $x_i$ cannot belong to $T^{(h)}$ for $k_h =1$ because it is assumed that $x_i$ does not form a message-connected $0$-cycle. Thus, we can conclude that $x_i$ is the same as being sent in the uncoded form from Lemma \ref{lemma:uncoded}.
\end{IEEEproof}
\label{lemma:uncoded2}
\end{lemma}

The following two theorems show some cases that side information is not critical.

\vspace{2mm}
\begin{theorem}
Assume that there is no message-connected $0$-cycle containing $x_b$ and a message with an index in $M_i\cap M_c$ for $a\in M_i$, $b\in (\cup_{j\in Z[|S|]} M_j)\setminus M_i$, and $i\in Z[|S|]$. Then, receivers wanting $x_a$ do not need to know $x_b$ as side information.
\begin{IEEEproof}
It is divided into two cases that $x_b$ does not form a $0$-cycle with a message with an index in $M_i\cap M_c$ or forms a message-disconnected $0$-cycle with a message with an index in $M_i\cap M_c$. 

For the first case, it is again classifed into two cases. The first one is that every message with an index in $M_i\cap M_c$ does not form a $0$-cycle. Then, from Lemma \ref{lemma:uncoded}, messages with indices in $M_i\cap M_c$ are the same as being sent in the uncoded forms. If $a\in M_i\cap M_c$, it is trivially proved. For $a\in M_i\setminus M_c$, knowing $x_b$ as side information does not help to recover $x_a$ because there is no path from $x_b$ to $x_a$ in the message graph excluding messages with indices in $M_i\cap M_c$, which means that the encoded messages related with $x_b$ do not have any information of $x_a$. The second one is that $x_b$ does not form a $0$-cycle. Then, $x_b$ is the same as being sent in the uncoded form and thus having $x_b$ as side information is not critical from Lemma \ref{lemma:uncoded}. 

For the second case, it can be proved by the similar method in the proof of Lemma \ref{lemma:uncoded2}. Assume that the maximum $0$-cycle including $x_b$ is represented as messages $T=\{x_b,x_{b_1},x_{b_2},\ldots,x_{b_t}\}$. If $x_a\notin T$, the theorem is easily proved because $x_a$ is the same as being sent in the uncoded form from Lemma \ref{lemma:uncoded} and thus we assume that $x_a\in T$. Then, partition $T$ into subsets of messages $T_1,T_2,\ldots,T_k$ as in Lemma \ref{lemma:uncoded2}. Assume that $T_1$ contains all messages related with $M_i$. If $x_b\notin T_1$, it is directly proved. If $x_b\in T_1$, it is similarly proved as in Lemma \ref{lemma:uncoded2}. Specifically, if we continue the same procedure as the above, $x_b$ does not belong to $T^{(j)}$ for some $j$ or $x_b$ belongs to $T^{(h)}$ such that $k_h=1$ or $T^{(h)}$ is $\phi$. However, $x_a$ and $x_b$ cannot belong to $T^{(h)}$ together because it is assumed that $x_b$ does not form a message-connected $0$-cycle with messages with indices in $M_i\cap M_c$. It is noted that there is no path between $x_a$ and $x_b$ not including messages with indices in $M_i\cap M_c$ in the message graph. Thus, it can be concluded that $x_a$ or $x_b$ is the same as being sent in the uncoded form and receivers wanting $x_a$ do not need to know $x_b$ as side information.
\end{IEEEproof}
\end{theorem}

\vspace{2mm}
\begin{theorem}
For $a\in M_i\setminus M_c$ and $i\in Z[|S|]$, assume that $R_k$ for $k\in Z[m]$ wants $x_a$. If every message in $\{x_q | q\in M_i\cap M_c\cap\mathcal{Y}_k\}$ does not form a message-connected $0$-cycle, side information of $R_k$ in $(\cup_{j\in Z[|S|]} M_j)\setminus M_i$ is not critical.
\begin{IEEEproof}
Since every message in $\{x_q | q\in M_i\cap M_c\cap\mathcal{Y}_k\}$ does not form a message-connected $0$-cycle, those messages are the same as being sent in the uncoded forms from Lemma \ref{lemma:uncoded2}. Thus, we can assume that $R_k$ knows all messages with indices in $M_i\cap M_c$ as side information. Considering the fitting matrix $F$ and linearly independent columns of $F$ as $F_1,\ldots,F_N$, $\Sigma_z b_z^k F_z=e_a+\Sigma_{y\in\mathcal{X}_{k}}c_y^ke_y$ for some $b_z^k, c_y^k\in \mathbb{F}_q$ from Claim \ref{claim:ICSI}. In fact, the $k$th column of $F^{(i)}$ is given as $e_a+\Sigma_{y\in\mathcal{X}_{k}}d^k_ye_y$ from the definition of the fitting matrix, where $d_y^k\in\mathbb{F}_q$. Thus, the $k$th column of $F^{(h)}$ for all $h\in Z[|S|]\setminus\{i\}$ can be deleted because the minimum rank of the modified fitting matrix is smaller than or equal to the minimum rank of the original one and $R_{k}$ can obtain $x_a$ from the $k$th column of $F^{(i)}$. It means that side information of $R_k$ in $(\cup_{j\in Z[|S|]} M_j)\setminus M_i$ is not critical.
\end{IEEEproof}
\label{theorem:common}
\end{theorem}

From Theorem \ref{theorem:common}, we have the following corollary. 

\vspace{2mm}
\begin{cor}
For $R_k$ such that $f(k)\in M_c$ and $k\in Z[m]$, assume that senders $s_{k_1},\ldots,s_{k_i}$ have $x_{f(k)}$ and $M_c^k$ is a set of message indices of those senders. If every message in $\{x_q | q\in M_c^k \cap M_c \cap \mathcal{Y}_k\}$ does not form a message-connected $0$-cycle, side information of $R_k$ in $(\cup_{j\in Z[|S|]} M_j)\setminus M_p$ for certain $p\in \{k_1,\cdots,k_i\}$ is not critical.
\begin{IEEEproof}
It is similar to the proof of Theorem \ref{theorem:common} and the only difference is that for the $k$th columns of submatrices of $F$, it is not determined which $k$th column of submatrices has the non-zero value in the $f(k)$th position. Thus, if the $k$th column of $F^{(p)}$ is selected to have the non-zero value in the $f(k)$th position, side information of $R_k$ in $(\cup_{j\in Z[|S|]} M_j)\setminus M_p$ is not critical as in Theorem \ref{theorem:common}.    
\end{IEEEproof}
\label{cor:commonpart}
\end{cor}


From Theorem \ref{theorem:common} and Corollary \ref{cor:commonpart}, it is noted that if all receivers satisfy the above conditions, an optimal $(\mathcal{G},M)$-IC can be obtained by the single sender IC problem with the modified side information graph because side information in the other senders is not critical. 

Existence of a $0$-cycle is a necessary and sufficient condition for reducing codelength by index coding in the single sender problem \cite{ICSIE}. In the multiple sender case, a message-connected $0$-cycle has the same property as shown in the following theorem.

\vspace{2mm}
\begin{theorem}
For a given side information graph $\mathcal{G}$ and a given message graph $U$, $N_{\rm opt}^q(\mathcal{G},M)=n$ if and only if there is no message-connected $0$-cycle in $\mathcal{G}$.
\begin{IEEEproof}
Necessity: Assume that there is a message-connected $0$-cycle represented by $\{x_1,x_2,\ldots,x_{n^\prime}\}$. Since there exists at least one path between any pair of message nodes corresponding to the $0$-cycle in $U$, there exists a spanning tree $\mathcal{T}$. Then, a $(\mathcal{G},M)$-IC with codelength $n-1$ can be constructed by using the index code $\{x_i+x_j : i {\rm~and~} j {\rm~are~connected~in~} \mathcal{T}\}$. From this index code with codelength $n^\prime-1$, we can get the sum or difference between any two messages $x_i, x_j$ for $i,j\in Z[n^\prime]$. Since every receiver wanting one of messages in $\{x_1,x_2,\ldots,x_{n^\prime}\}$ has at least one side information symbol in $\{x_1,x_2,\ldots,x_{n^\prime}\}$, it can recover the wanted message. If we send the remaining messages in $\{x_{n^\prime+1},\ldots,x_n\}$ as the uncoded forms, every receiver in $\mathcal{G}$ can obtain what it wants and the codelength is $n-1$.

Sufficiency: If there is no message-connected $0$-cycle, every message does not form a message-connected $0$-cycle. Then, from Lemma \ref{lemma:uncoded2}, every message is the same as being sent in the uncoded form.
\end{IEEEproof}
\label{theorem:cycle}
\end{theorem}

\vspace{2mm}
\begin{remark}
For $m=n$, existence of a message-connected cycle in \cite{TSIC} is not a necessary and sufficient condition for reducing codelength. For example, assume that $M_1=\{1,4\}$, $M_2=\{2,3,4\}$, and $f(i)=i$ for $i\in Z[4]$. Let $\mathcal{X}_1=\{2\}$, $\mathcal{X}_2=\{3\}$, $\mathcal{X}_3=\{1\}$, and $\mathcal{X}_4=\{2\}$. Then, there is no message-connected cycle. However, the entire graph is a message-connected $0$-cycle and we can make the index code $(x_1+x_4, x_2+x_4, x_2+x_3)$ with codelength $3$.
\end{remark}

\vspace{2pt}
\section{Extension to Cellular Network}\label{sec:cellularnetwork}

Now, we consider the index coding in a cellular network scenario, where receivers can receive a subset of sub-codewords because some senders cannot cover all receivers. If each receiver belongs to coverage of only one sender, it just reduces to disjoint single sender index coding problems. However, there is a possibility to reduce index codelength more efficiently if some receivers belong to coverage of more than one sender. We first describe index coding in a cellular network as follows.

\subsection{Problem Description: Two Sender Case}

In the linear index coding with a cellular network, the problem setting is almost identical to that of the linear index coding problem for the multiple sender case except that some receivers are restricted to receive a subset of sub-codewords. For simplicity, we assume that $m=n$, $f(i)=i$ for $i\in Z[m]$, $|S|=2$, and the field size $q=2$. 

Then, $\mathcal{G}$ is a unipartite side information graph, where each node represents both a receiver and a message \cite{ICSI}. A directed edge from a node (say 1) to another node (say 2) means that receiver 1 has message 2 as side information.

Since a cellular network is assumed, there are three types of receivers based on coverage of senders. Let $R(s_j)=\{i | R_i {\rm ~can ~only ~receive ~the~sub}$-${\rm codeword ~from~} s_j \}$ for $j\in Z[2]$, $R(s_c)=\{i| R_i {\rm ~can ~receive ~the~entire ~codeword}\}$, and $R=\{R(s_1),R(s_2),R(s_c)\}$. 

Then, a linear index code for a cellular network is defined as follows.

\vspace{2mm}
\begin{definition}
A linear index code over $\mathbb{F}_2$ for a cellular network with two senders, denoted by a $(\mathcal{G},M,R)$-IC is a set of codewords having:
\begin{enumerate}
\item Generator submatrices $G^{(i)}\in\mathbb{F}_2^{|M_i|\times N_i}$ for $i\in Z[2]$, where $N_i$ denotes the codelength of the sub-codeword $\bold{x}_{M_i}G^{(i)}$ generated by $s_i$.
\item Decoding functions $D_j$ satisfying 
\begin{equation}
D_j(\bold{x}_{M_1}G^{(1)},\bold{x}_{\mathcal{X}_j})=x_{f(j)} {\rm ~if~} j\in R(s_1)
\nonumber
\end{equation}
\begin{equation}
D_j(\bold{x}_{M_2}G^{(2)},\bold{x}_{\mathcal{X}_j})=x_{f(j)} {\rm ~if~} j\in R(s_2)
\nonumber
\end{equation}
\begin{equation}
D_j(\bold{x}_{M_1}G^{(1)},\bold{x}_{M_{2}}G^{(2)},\bold{x}_{\mathcal{X}_j})=x_{f(j)} {\rm ~if~} j\in R(s_c)
\nonumber
\end{equation}
for all $j\in Z[m]$, $\bold{x}\in\mathbb{F}_2^n$.
\end{enumerate}
\end{definition}
\vspace{2mm}

It is noted that an $n\times N$ generator matrix $G$ for a $(\mathcal{G},M,R)$-IC is constructed by using generator submatrices $G^{(i)}$ for $i\in Z[2]$ and the codelength $N$ is $N_1+N_{2}$. Let $N_{\rm opt}^2(\mathcal{G},M,R)$ be the optimal codelength of a $(\mathcal{G},M,R)$-IC. 

Since receivers with indices in $R(s_i)$ cannot receive the sub-codeword from $s_j$ for $i,j\in Z[2]$ with $i\neq j$, we have the following proposition.

\vspace{2mm}
\begin{proposition}
For a receiver with an index in $R(s_i)$, side information in $M_j\setminus M_i$ is not critical for $i,j\in Z[2]$ with $i\neq j$.
\begin{IEEEproof}
Since every receiver with an index in $R(s_i)$ cannot receive encoded messages made from using messages with indices in $M_j\setminus M_i$, those receivers can recover their wanted messages without using side information in $M_j\setminus M_i$.
\end{IEEEproof}
\label{prop:cellularsideinfo}
\end{proposition}
\vspace{2mm}

Thus, we can assume that every receiver with an index in $R(s_i)$ has side information only in $M_i$ for $i\in Z[2]$. 


\subsection{Encoding Procedure of $(\mathcal{G},M,R)$-IC}

Now, we first introduce a fitting matrix for the cellular network case, which is similar to that of the multiple sender case. Since some receivers receive a subset of sub-codewords, a new parameter instead of the minimum rank of the fitting matrix is needed to represent codelength. The fitting matrix for the cellular network consists of three submatrices as shown in the following definition.

\vspace{2mm}
\begin{definition}
The fitting matrix $F$ of size $n\times (|R(s_1)|+|R(s_2)|+2|R(s_c)|)$ for a $(\mathcal{G},M,R)$-IC is described as follows:
\begin{enumerate}
\item There are three submatrices $F^{(1)},F^{(2)},F^{(3)}$ corresponding to three types of receivers $R(s_1)$, $R(s_2)$, and $R(s_c)$, respectively.
\item For $F^{(i)}$, each column represents each receiver and it is the same as the transpose of the row of the conventional fitting matrix of the single sender IC problem \cite{ICSI}, where $i\in Z[2]$ and $F^{(i)}$ is used for encoding in $s_i$. 
\item For $F^{(3)}$, each receiver is represented by two columns used for encoding in $s_1$ and $s_2$.
\item $F_{i,k_j}^{(3)}=0$ for $i\notin M_j$, $j\in Z[2]$, and $k\in Z[m]$.
\item $\Sigma_{j=1}^{2}F_{k,k_j}^{(3)}=1$ for $k\in Z[m]$.
\item $\Sigma_{j=1}^{2}F_{i_1,k_j}^{(3)}=0$ for $i_1\in\mathcal{Y}_k$ and $k\in Z[m]$.
\item For $i_2\in M_j\cap\mathcal{X}_k$, $j\in Z[2]$, and $k\in Z[m]$, $F_{i_2,k_j}^{(3)}$ is any element of $\mathbb{F}_2$.

\end{enumerate}
\label{def:cellularfitting}
\end{definition}
\vspace{2mm}

Let $V_i$ be a vector space spanned by columns of $F^{(i)}$ for $i\in Z[3]$. Unlike the multiple sender case, the optimal codelength of linear index codes in the cellular network is not the minimum rank of the fitting matrices. The following theorem shows the optimal codelength of linear index codes in the cellular network.

\vspace{2mm}
\begin{theorem}
$N_{\rm opt}^2(\mathcal{G},M,R)$ is the minimum of ${\rm dim}(V_1+V_2+V_3)+{\rm dim}(V_1\cap V_2)$.
\begin{IEEEproof}
First, it is easily understood that the fitting matrix for the cellular network with two senders can be a generator matrix of a $(\mathcal{G},M,R)$-IC. From the fitting matrix, the codelength of the index code can be given as ${\rm dim}(V_1)+{\rm dim}(V_2)+{\rm dim}(V_3)-{\rm dim}(V_3\cap(V_1+V_2))$ because receivers with indices in $R(s_i)$ have to receive encoded messages by $F^{(i)}$ with codelength ${\rm dim}(V_i)$ for $i\in Z[2]$ and receivers with indices in $R(s_c)$ can recover their wanted messages by additionally receiving independent messages (columns of $F^{(3)}$) from $V_1+V_2$ with codelength ${\rm dim}(V_3)-{\rm dim}(V_3\cap (V_1+V_2))$. Since ${\rm dim}(V_1+V_2+V_3)={\rm dim}(V_1+V_2)+{\rm dim}(V_3)-{\rm dim}(V_3\cap (V_1+V_2))$ from the inclusion-exclusion principle, codelength can be represented as ${\rm dim}(V_1+V_2+V_3)-{\rm dim}(V_1+V_2)+{\rm dim}(V_1)+{\rm dim}(V_2)={\rm dim}(V_1+V_2+V_3)+{\rm dim}(V_1\cap V_2)$.

Assume that $G$ is a generator matrix of a $(\mathcal{G},M,R)$-IC. Then, it can be partitioned into two parts based on encoding of each sender. Let $G_v^{(i)}$ be a vector space spanned by columns of $G^{(i)}$ for $i\in Z[2]$. Then, the codelength of this index code is ${\rm dim}(G_v^{(1)})+{\rm dim}(G_v^{(2)})$. From the similar methods in Claim \ref{claim:ICSI} and the proof of Theorem \ref{theorem:multisenderencoding}, it is noted that the fitting matrix for the cellular network can be made from columns of $G$. Specifically, a vector space $V_i$ can be made from $G_v^{(i)}$ for $i\in Z[2]$ and $V_3$ can be made from $G_v^{(1)}$ and $G_v^{(2)}$. Similar to Corollary \ref{cor:reduce}, we only consider cases satisfying 5), 6) of Definition \ref{def:cellularfitting} because we can consider columns with zero coefficients as all-zero columns in the perspective of the optimal codelength.

Then, it is noted that ${\rm dim}(V_1+V_2+V_3)\leq {\rm dim}(G_v^{(1)}+G_v^{(2)})$ and ${\rm dim}(V_1\cap V_2)\leq {\rm dim}(G_v^{(1)}\cap G_v^{(2)})$. Thus, ${\rm dim}(V_1+V_2+V_3)+{\rm dim}(V_1\cap V_2)\leq {\rm dim}(G_v^{(1)}+G_v^{(2)})+{\rm dim}(G_v^{(1)}\cap G_v^{(2)})={\rm dim}(G_v^{(1)})+{\rm dim}(G_v^{(2)})$, which means that the optimal codelength is the minimum of ${\rm dim}(V_1+V_2+V_3)+{\rm dim}(V_1\cap V_2)$.

\end{IEEEproof}
\label{theorem:cellularencoding}
\end{theorem}
\vspace{2mm}


The example of Proposition \ref{prop:cellularsideinfo} and the fitting matrix for the cellular network is given as follows.

\begin{figure}[t]
\centering
\subfigure[]{\includegraphics[scale=0.5]{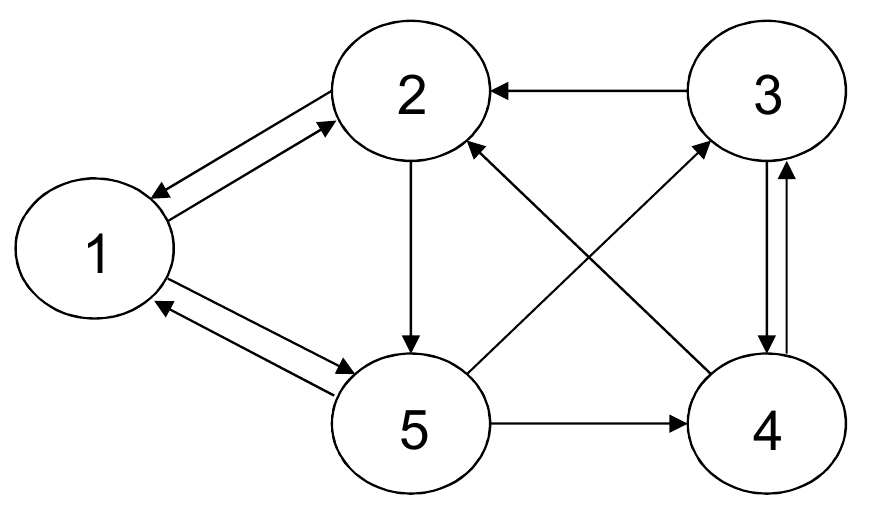}
\label{fig:cellularfittingex}}
\subfigure[]{\includegraphics[scale=0.5]{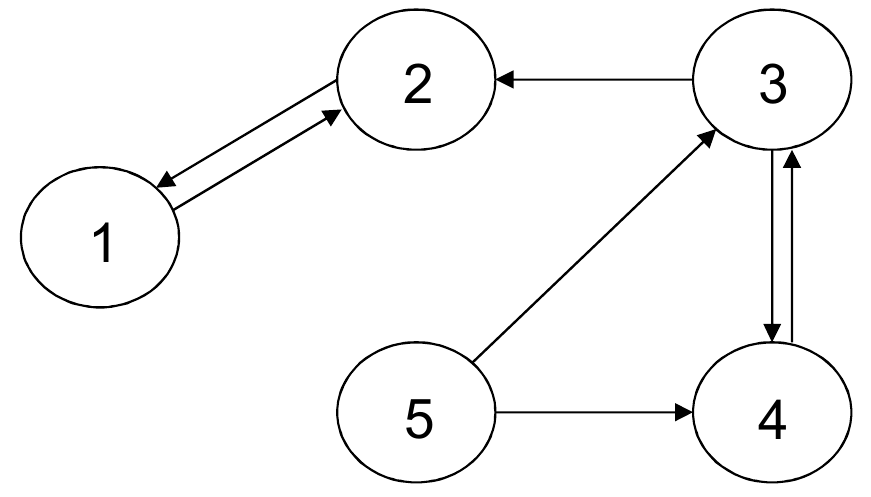}
\label{fig:cellularfittingex1}}
\caption{A side information graph $\mathcal{G}$ of Example \ref{ex:cellularfitting}: (a) A side information graph $\mathcal{G}$. (b) The modified side information graph $\hat{\mathcal{G}}$ from Proposition \ref{prop:cellularsideinfo}.} 
\end{figure}

\vspace{2mm}
\begin{example}
Let $m=5$, $M=\{M_1,M_2\}$, $M_1=\{1,2,3\}$, $M_2=\{3,4,5\}$, $R(s_1)=\{1,2\}$, $R(s_2)=\{4,5\}$, and $R(s_c)=\{3\}$. A side information graph $\mathcal{G}$ is given in Fig. \ref{fig:cellularfittingex}. From Proposition \ref{prop:cellularsideinfo}, $\mathcal{G}$ can be transformed into $\hat{\mathcal{G}}$ as in Fig. \ref{fig:cellularfittingex1}. Then, the fitting matrix of a $(\hat{\mathcal{G}},M,R)$-IC is derived as

\begin{equation}
F=\begin{pmatrix}F^{(1)}&F^{(2)}&F^{(3)}\end{pmatrix},
\nonumber
\end{equation} where

\begin{equation}
F^{(1)}=\begin{pmatrix}1&*\\{*}&1\\0&0\\0&0\\0&0\end{pmatrix},  F^{(2)}=\begin{pmatrix}0&0\\{0}&0\\{*}&{*}\\1&*\\{0}&1\end{pmatrix}, F^{(3)}=\begin{pmatrix}0&0\\{*}&0\\{c}&{1+c}\\0&*\\{0}&0\end{pmatrix}
\nonumber
\end{equation}
and ${*}$ denotes any element of $\mathbb{F}_q$ and $c\in\mathbb{F}_2$. By finding a fitting matrix with the minimum value of ${\rm dim}(V_1+V_2+V_3)+{\rm dim}(V_1\cap V_2)$, we can find an optimal generator matrix.

\label{ex:cellularfitting}
\end{example}
\vspace{2mm}

As mentioned before, we only consider a side information graph $\hat{\mathcal{G}}$ by deleting uncritical side information from Proposition \ref{prop:cellularsideinfo} as in Example \ref{ex:cellularfitting}.

\subsection{Properties of $(\mathcal{G},M,R)$-IC}

In this section, we study some properties of a $(\mathcal{G},M,R)$-IC for a cellular network with two senders. First, we have the following definition describing a new type of a side information graph.

\vspace{2mm}
\begin{definition}
For the cellular network, a subgraph $\mathcal{G}^\prime$ of $\mathcal{G}$ is denoted by $\mathcal{H}$ if every message is side information of at least one receiver in $\mathcal{G}^\prime$.  
\end{definition}
\vspace{2mm}

There is an example for $\mathcal{H}$.

\vspace{2mm}
\begin{example}
Assume the same situation as in Example \ref{ex:cellularfitting}. Then, $\hat{\mathcal{G}}$ is not $\mathcal{H}$ because no receiver has $x_5$ as side information. However, a subgraph induced by nodes $\{1,2,3,4\}$ is $\mathcal{H}$ because every message is side information of at least one receiver.
\end{example}
\vspace{2mm}

The following propositions show some relationships between a $(\mathcal{G},M)$-IC and a $(\mathcal{G},M,R)$-IC.

\vspace{2mm}
\begin{proposition}
$N_{\rm opt}^2(\mathcal{G},M,R)=N_{\rm opt}^2(\mathcal{G},M)$ if ${\rm dim}(V_1\cap V_2)=0$ holds for any fitting matrix of a $(\mathcal{G},M,R)$-IC.
\begin{IEEEproof}
From Proposition \ref{prop:sideinformation} and Definitions \ref{def:fittingmatrix} and \ref{def:cellularfitting}, it is noted that the fitting matrix for a $(\mathcal{G},M)$-IC and the fitting matrix for a $(\mathcal{G},M,R)$-IC have the same form except for column permutation and all-zero columns. Since $N_{\rm opt}^2(\mathcal{G},M,R)$ is the minimum value of ${\rm dim}(V_1+V_2+V_3)+{\rm dim}(V_1\cap V_2)$ and $N_{\rm opt}^2(\mathcal{G},M)$ is the minimum value of ${\rm dim}(V_1+V_2+V_3)$, they are the same if ${\rm dim}(V_1\cap V_2)=0$.
\end{IEEEproof}
\label{prop:optimalcodelength}
\end{proposition}
\vspace{2mm}

\vspace{2mm}
\begin{proposition}
A fitting matrix with ${\rm dim}(V_1\cap V_2)\neq0$ exists if and only if $\mathcal{H}$ satisfying one of the followings exists, where $\mathcal{H}$ consists of receivers with indices in $R(s_1)\cup R(s_2)$ and contains at least one receiver with an index in $R(s_i)$ for all $i\in Z[2]$.
\begin{enumerate}
\item A receiver with an index in $R(s_1)$ has a message with an index in $R(s_2)$ as side information or vice versa. 
\item A receiver with an index in $R(s_1)$ and a receiver with an index in $R(s_2)$ have the same message as side information.
\end{enumerate}
\begin{IEEEproof}
From the combinatoric point of view, a fitting matrix with ${\rm dim}(V_1\cap V_2)\neq0$ exists if and only if there exist column vectors such that the sum of $a$ column vectors of $F^{(1)}$ and $b$ column vectors of $F^{(2)}$ is the all-zero vector, where $a,b\neq 0$ and the sum of those $a$ column vectors is not the all-zero vector. First, it is easily noted that there exist column vectors such that the sum of $a$ column vectors of $F^{(1)}$ and $b$ column vectors of $F^{(2)}$ is the all-zero vector, where $a,b\neq 0$ if and only if those $a+b$ receivers (messages) form $\mathcal{H}$. It is because each message has to be known to at least one receiver in order for the element corresponding to each message in the sum of those vectors to be $0$.
%
%
%

Next, we deal with the additional condition that there exists a case that the sum of those $a$ column vectors is not the all-zero vector. It is noted that this condition holds if and only if a column corresponding to one of $a$ receivers and a column corresponding to one of $b$ receivers can have $1$ in the same position, which reduces to the above two cases. 
\end{IEEEproof}
\label{prop:fittingcases}
\end{proposition}
\vspace{2mm}

There is an example for Propositions \ref{prop:optimalcodelength} and \ref{prop:fittingcases}.

\begin{figure}[t]
\centering
\includegraphics[scale=0.6]{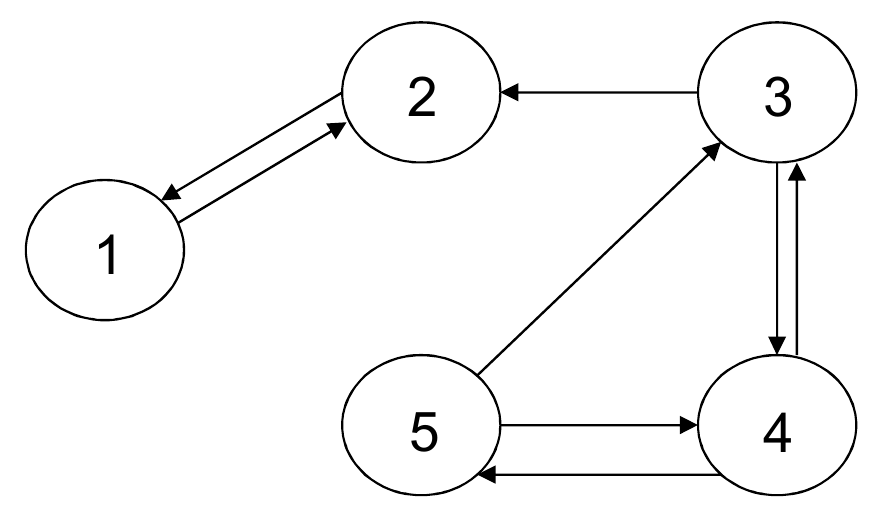}
\caption{A side information graph $\mathcal{G}$ of Example \ref{ex:propexcellular}.} 
\label{fig:propexcellular}
\end{figure}

\vspace{2mm}
\begin{example}
Let $m=5$, $M=\{M_1,M_2\}$, $M_1=\{1,2,3\}$, $M_2=\{3,4,5\}$, $R(s_1)=\{1,2\}$, $R(s_2)=\{4,5\}$, and $R(s_c)=\{3\}$. A side information graph $\mathcal{G}$ is given in Fig. \ref{fig:propexcellular}. Although a subgraph induced by $\{1,2,4,5\}$ forms $\mathcal{H}$, 1) and 2) of Proposition \ref{prop:fittingcases} do not hold. Thus, ${\rm dim}(V_1\cap V_2)=0$ always holds from Proposition \ref{prop:fittingcases} and $N_{\rm opt}^2({\mathcal{G}},M,R)=N_{\rm opt}^2({\mathcal{G}},M)$ from Proposition \ref{prop:optimalcodelength}. 
\label{ex:propexcellular}
\end{example}
\vspace{2mm}

A cycle in a side information graph $\mathcal{G}$ is known to be important in the single sender index coding problem and there are lots of cycle-cover algorithms to find the suboptimal index codelength. Along with these, we classify which cycle can be used to reduce index codelength in the cellular network.

\vspace{2mm}
\begin{theorem}
In the cellular network, cycles can be classified into the following cases based on a possibility of reducing codelength.
\begin{enumerate}
\item Index codelength cannot be reduced from a message-disconnected cycle and a message-connected cycle which consists of receivers with indices in $R(s_1)\cup R(s_2)$ and contains at least one receiver with an index in $R(s_i)$ for all $i\in Z[2]$.
\item A message-connected cycle consisting of receivers with indices in $R(s_i)$ for $i\in Z[2]$ and a message-connected cycle containing at least one receiver with an index in $R(s_c)$ can reduce index codelength.
\end{enumerate}
\begin{IEEEproof}
Since a message-disconnected cycle is a message-disconnected $0$-cycle and $N_{\rm opt}^2(\mathcal{G},M,R)\geq N_{\rm opt}^2(\mathcal{G},M)$, a message-disconnected cycle cannot reduce codelength from Theorem \ref{theorem:cycle}. Next, for a message-connected cycle described in 1), assume that $a$ receivers with indices in $R(s_1)$ and $b$ receivers with indices in $R(s_2)$ form a message-connected cycle. Since each receiver can receive encoded messages from only one sender and two subgraphs induced by $a$ receivers and $b$ receivers, respectively are acyclic, $a+b$ transmissions are required.

It is trivial that a message-connected cycle consisting of receivers with indices in $R(s_i)$ for $i\in Z[2]$ can reduce codelength. For a message-connected cycle containing at least one receiver with an index in $R(s_c)$, it is noted that every receiver has one message as side information and one message cannot be side information of two receivers. Since each receiver with an index in $R(s_i)$ for $i\in Z[2]$ of that cycle has side information with an index in $M_i$, each receiver can recover the wanted message if $s_i$ transmits the sum of the wanted message and its side information.
 
Similar to the perspective of a message graph in Theorem \ref{theorem:cycle}, we make a graph consisting of the messages. We add an undirected edge between the wanted message and its side information for each receiver with an index in $R(s_1)\cup R(s_2)$, which represents each transmission of the sum of the wanted message and its side information. Then, there is no cycle in the resulting graph consisting of those receivers with indices in $R(s_1)\cup R(s_2)$ because there is no cycle in the subgraph of $\mathcal{G}$ induced by those receivers.

Since there is no cycle in the resulting graph and there is at least one message with an index in $M_c$ due to message-connectivity, we can make a spanning tree including the other messages with indices in $R(s_c)$ while maintaining the above undirected edges. Then, new connected edges correspond to transmissions of the sum of two messages as same as before by a sender capable of encoding those messages. Then, every receiver can recover its wanted message because receivers with indices in $R(s_c)$ can receive all encoded messages corresponding to edges in the spanning tree.
\end{IEEEproof}
\label{theorem:cyclecasescellular}
\end{theorem}
\vspace{2mm}

Next, we study properties of $N_{\rm opt}^2(\mathcal{G},M,R)$ in the cellular network. In Section \ref{sec:Properties}, Theorem \ref{theorem:cycle} shows that existence of a message-connected $0$-cycle is a necessary and sufficient condition for reducing codelength in the index coding problem with multiple senders. However, it does not hold for the cellular network. Furthermore, a receiver with no side information can help to reduce codelength in the cellular network as shown in the following example.

\vspace{2mm}
\begin{example}
Let $M_1=\{1,2,3\}$, $M_2=\{1,3\}$, $R(s_1)=\{1\}$, $R(s_2)=\{3\}$, $R(s_c)=\{2\}$, $\mathcal{X}_1=\{2,3\}$, $\mathcal{X}_2=\phi$, and $\mathcal{X}_3=\{1\}$. Although a subgraph of $\mathcal{G}$ induced by $\{1,3\}$ is a message-connected $0$-cycle, it cannot reduce codelength from Theorem \ref{theorem:cyclecasescellular} and thus the optimal codelength for this subgraph is $2$. Now, consider $R_2$ which has no side information. In the index coding problem with multiple senders or single sender, adding a receiver with no side information increases the optimal codelength by one when $m=n$. However, in this example, every receiver can recover its wanted message if $s_1$ transmits $(x_1+x_2+x_3)$ and $s_2$ transmits $(x_1+x_3)$. From this example, it is noted that the general condition for $N_{\rm opt}^2(\mathcal{G},M,R)=n$ is hard to be found through the conventional properties of index coding.

\label{ex:reducingcodelengthcellular}
\end{example}
\vspace{2mm}

Now, we discuss some properties of $N_{\rm opt}^2(\mathcal{G},M,R)$.

\vspace{2mm}
\begin{proposition}
Assume that ${\rm dim}(V_1\cap V_2)=0$ always holds. Then, $N_{\rm opt}^2(\mathcal{G},M,R)=n$ if and only if there is no messge-connected $0$-cycle in $\mathcal{G}$.
\begin{IEEEproof}
From Proposition \ref{prop:optimalcodelength} and Theorem \ref{theorem:cycle}, it is easily proved.
\end{IEEEproof}
\end{proposition}
\vspace{2mm}

Similar to Proposition \ref{prop:fittingcases}, we discuss the condition for ${\rm dim}(V_3\cap (V_1+V_2))\neq 0$. If ${\rm dim}(V_3\cap (V_1+V_2))\neq 0$, there are some column vectors in the fitting matrix such that the sum of them is the all-zero vector, where $u$ columns of them are in $F^{(3)}$. In the following observation, we classify those $u$ columns in $F^{(3)}$ into some receivers or certain vectors.  

\vspace{2mm}
\begin{observation}
Using notations in Definition \ref{def:cellularfitting}, $u$ columns in $F^{(3)}$ correspond to some receivers or certain vectors based on relationships between selected columns in $F^{(3)}$.

\begin{enumerate}
\item The cases that some columns correspond to some receivers:
      \begin{enumerate}
      \item If both the $k_1$th column and the $k_2$th column are selected for $k\in Z[m]$, the sum of them corresponds to $R_k$.
      \item If only the $k_i$th column is selected among the $k_1$th column and the $k_2$th column for $i\in Z[2]$ and $k\in M_i\setminus M_c$, the $k_i$th column corresponds to the receiver whose side information corresponds to $M_c\cup(\mathcal{X}_k\cap M_i)$ and wanted message is $x_k$.
      \end{enumerate}
\item The case that some columns correspond to certain vectors:
		\begin{enumerate}
		\item If only the $k_j$th column is selected among the $k_1$th column and the $k_2$th column for $i,j\in Z[2]$ such that $i\neq j$ and $k\in M_i$, the $k_j$th column is any column with $0$s in the positions in $Z[n]\setminus(M_c\cup(\mathcal{X}_k\cap M_j))$.
		\item It is said that a message (say $x_d$) is covered by the above columns corresponding to certain vectors if one of those columns can have $1$ at the $d$th position.
		\end{enumerate}
\end{enumerate}

%

%
\label{ob:columncase}
\end{observation}
\vspace{2mm}

The following observation shows a necessary and sufficient condition for ${\rm dim}(V_3\cap (V_1+V_2))\neq 0$ from the combinatoric point of view.

\vspace{2mm}
\begin{observation}
A fitting matrix with ${\rm dim}(V_3\cap (V_1+V_2))\neq 0$ exists if and only if there are $a$ receivers with indices in $R(s_1)\cup R(s_2)$, $b$ receivers corresponding to receivers in Observation \ref{ob:columncase}, and $b^\prime$ columns corresponding to certain vectors in Observation \ref{ob:columncase} for $a\neq 0$ and $b+b^\prime\neq 0$ satisfying one of the following nine cases:

\begin{enumerate}
\item The cases for $b=0$:
\begin{enumerate}
\item Each of $a$ messages is known to at least one of $a$ receivers or is covered by $b^\prime$ columns. At this point, at least one of $a$ messages has to be covered by $b^\prime$ columns.
\item $a$ receivers (messages) form $\mathcal{H}$ and none of $a$ messages is covered by $b^\prime$ columns. Let one of $a$ receivers have $x_d$ as side information such that $x_d$ does not belong to $a$ messages. Then, $x_d$ is covered by $b^\prime$ columns.
\end{enumerate}
\item The cases for $b^\prime=0$ and $a+b$ receivers (messages) form $\mathcal{H}$:
\begin{enumerate}
      \item One of $a$ messages is side information of one of $b$ receivers.
      \item One of $b$ messages is side information of one of $a$ receivers.
      \item One of $a$ receivers and one of $b$ receivers have the same message as side information.
      \end{enumerate}
\item The cases for $b,b^\prime\neq0$ which are not reduced to the above cases:
\begin{enumerate}
\item $a+b$ receivers (messages) do not form $\mathcal{H}$. Also, all of $a$ messages are known to at least one of $a+b$ receivers and there is at least one of $b$ messages not known to $a+b$ receivers. In this case, those unknown messages are covered by $b^\prime$ columns.
\begin{enumerate}
      \item One of $a$ messages is side information of one of $b$ receivers.
      \item One of $b$ messages is side information of one of $a$ receivers.
      \item One of $a$ receivers and one of $b$ receivers have the same message as side information.
      \item $x_d$ is covered by $b^\prime$ columns and one of $a$ receivers has $x_d$ as side information, where $x_d$ does not belong to $a+b$ messages.
      \end{enumerate}
\end{enumerate}
\end{enumerate}

\label{ob:uncodedcellularcase}
\end{observation}
\vspace{2mm}

The following proposition shows that the cellular network case can be reduced to disjoint index coding problems if ${\rm dim}(V_3\cap(V_1+V_2))=0$ always holds.

\vspace{2mm}
\begin{proposition}
If ${\rm dim}(V_3\cap(V_1+V_2))=0$ always holds, $N_{\rm opt}^2(\mathcal{G},M,R)$ is the sum of the optimal codelength of subproblems induced by $R(s_1)$, $R(s_2)$, and $R(s_c)$.
\begin{IEEEproof}
Since ${\rm dim}(V_3\cap(V_1+V_2))=0$ always holds, codelength can be represented as ${\rm dim}(V_1)+{\rm dim}(V_2)+{\rm dim}(V_3)$.
\end{IEEEproof}
\label{prop:last}
\end{proposition}
\vspace{2mm}

From Proposition \ref{prop:last}, it is noted that an $R(s_i)$ induced subgraph does not have a message-connected $0$-cycle for $i\in Z[2]$ and an $R(s_c)$ induced subgraph does not have a message connected $0$-cycle if and only if $N_{\rm opt}^2(\mathcal{G},M,R)=n$ given that ${\rm dim}(V_3\cap(V_1+V_2))=0$ always holds. 

\vspace{2pt}
\section{Conclusions}\label{Sec:Conclusion}

In this paper, we studied linear index codes with multiple senders. The fitting matrix for a $(\mathcal{G},M)$-IC was introduced and we showed that the fitting matrix having the minimum rank can be an optimal generator matrix. Some properties of a $(\mathcal{G},M)$-IC based on a $0$-cycle were studied and whether given side information is critical or not was studied for multiple senders.

Furthermore, another index coding scenario was also considered, where each receiver can receive a subset of sub-codewords as in the cellular network. Another type of fitting matrices was introduced and it was proved that the optimal codelength of linear index codes can be found from these fitting matrices by minimizing the value of ${\rm dim}(V_1+V_2+V_3)+{\rm dim}(V_1\cap V_2)$. Finally, some properties on the optimal codelength for the cellular network were studied.

\end{document}